\documentclass[a4paper,twocolumn,superscriptaddress,11pt,accepted=2018-06-01]{quantumarticle}
\usepackage{graphicx,amsmath,amssymb,color}
\usepackage[normalem]{ulem}
\usepackage{amsfonts}
\usepackage[breaklinks,ocgcolorlinks,colorlinks=true,linkcolor=blue,citecolor=red]{hyperref}
\usepackage{latexsym}
\usepackage{amsfonts}
\usepackage{color,verbatim}

\usepackage[sort&compress,numbers]{natbib}

\newcommand{\ba}{\begin{eqnarray}}
\newcommand{\be}{\begin{equation}}
\newcommand{\ee}{\end{equation}}

\newcommand{\ea}{\end{eqnarray}}
\newcommand{\ban}{\begin{eqnarray*}}
\newcommand{\ean}{\end{eqnarray*}}
\newcommand{\Tr}{\operatorname{Tr}}

\newcommand{\ket}[1]{|#1\rangle}
\newcommand{\bra}[1]{\langle#1|}

\newcommand{\figref}[1]{Fig.~\ref{#1}}
\newcommand{\secref}[1]{Sec.~\ref{#1}}
\newcommand{\appref}[1]{App.~\ref{#1}}

\begin{document}
	
\title{Heralded generation of maximal entanglement in any dimension via incoherent coupling to thermal baths}

\author{Armin Tavakoli}
\affiliation{Department of Applied Physics, University of Geneva, 1211 Geneva, Switzerland}
\author{G\'eraldine Haack}
\affiliation{Department of Applied Physics, University of Geneva, 1211 Geneva, Switzerland}
\orcid{0000-0001-7893-0177}
\author{Marcus Huber}
\affiliation{Institute for Quantum Optics and Quantum Information (IQOQI), Austrian Academy of Sciences, Boltzmanngasse 3, A-1090 Vienna, Austria}
\orcid{0000-0003-1985-4623}
\author{Nicolas Brunner}
\affiliation{Department of Applied Physics, University of Geneva, 1211 Geneva, Switzerland}
\author{Jonatan Bohr Brask}
\affiliation{Department of Applied Physics, University of Geneva, 1211 Geneva, Switzerland}
\orcid{0000-0003-3859-0272}

\date{\today}

\begin{abstract}
We present a scheme for dissipatively generating maximal entanglement in a heralded manner. Our setup requires incoherent interactions with two thermal baths at different temperatures, but no source of work or control. A pair of $(d+1)$-dimensional quantum systems is first driven to an entangled steady state by the temperature gradient, and maximal entanglement in dimension $d$ can then be heralded via local filters. We discuss experimental prospects considering an implementation in superconducting systems.
\end{abstract}
	
\maketitle


\section{Introduction}
Entanglement is a key phenomenon distinguishing quantum from classical physics, and is the paradigmatic resource enabling many applications of quantum information science. Generating and maintaining entanglement is therefore a central challenge. Decoherence caused by unavoidable interactions of a system with its environment generally degrades entanglement, and significant effort is invested in minimising the effect of such dissipation in experiments. 

However, dissipation can also be advantageous, and may indeed be exploited for the generation of entangled quantum states under the right conditions \cite{Plenio1999,Kim2002,Jakobczyk2002,Braun2002,Benatti2003,Burgarth2007,Bellomo2008,Manzano2012}. In particular, it is possible for dissipative processes to drive the system into an entangled steady state \cite{Diehl2008,Verstraete2009,Kraus2008,Ticozzi2014,Tacchino2018}. This was studied in a variety of physical systems \cite{Schneider2002,Kastoryano2011,Wang2010,Reiter2013,Schuetz2013,Cai2010,Walter2013} and demonstrated experimentally for atomic ensembles \cite{Krauter2011}, trapped ions \cite{Barreiro2011,Lin2013}, and superconducting qubits \cite{Shankar2013}. The main ingredients are engineered decay processes and quantum bath engineering \cite{Vacanti2009,Reiter2012,Aron2014}, and coherent external driving is employed, which, from a thermodynamic point of view, can be considered a source of work.

More generally, it is natural to look for the minimal setting in which dissipative entanglement generation is possible. In particular, one may ask if entanglement can be generated from purely thermal processes alone, without the need for work input or external control. This can in principle be achieved in equilibrium situations, as any entangled state can be obtained as the ground state of a specific Hamiltonian. However, this requires highly nonlocal Hamiltonians which may be extremely difficult to implement in practice. 

\begin{figure}[t]
\begin{center}
\includegraphics[width=0.95\linewidth]{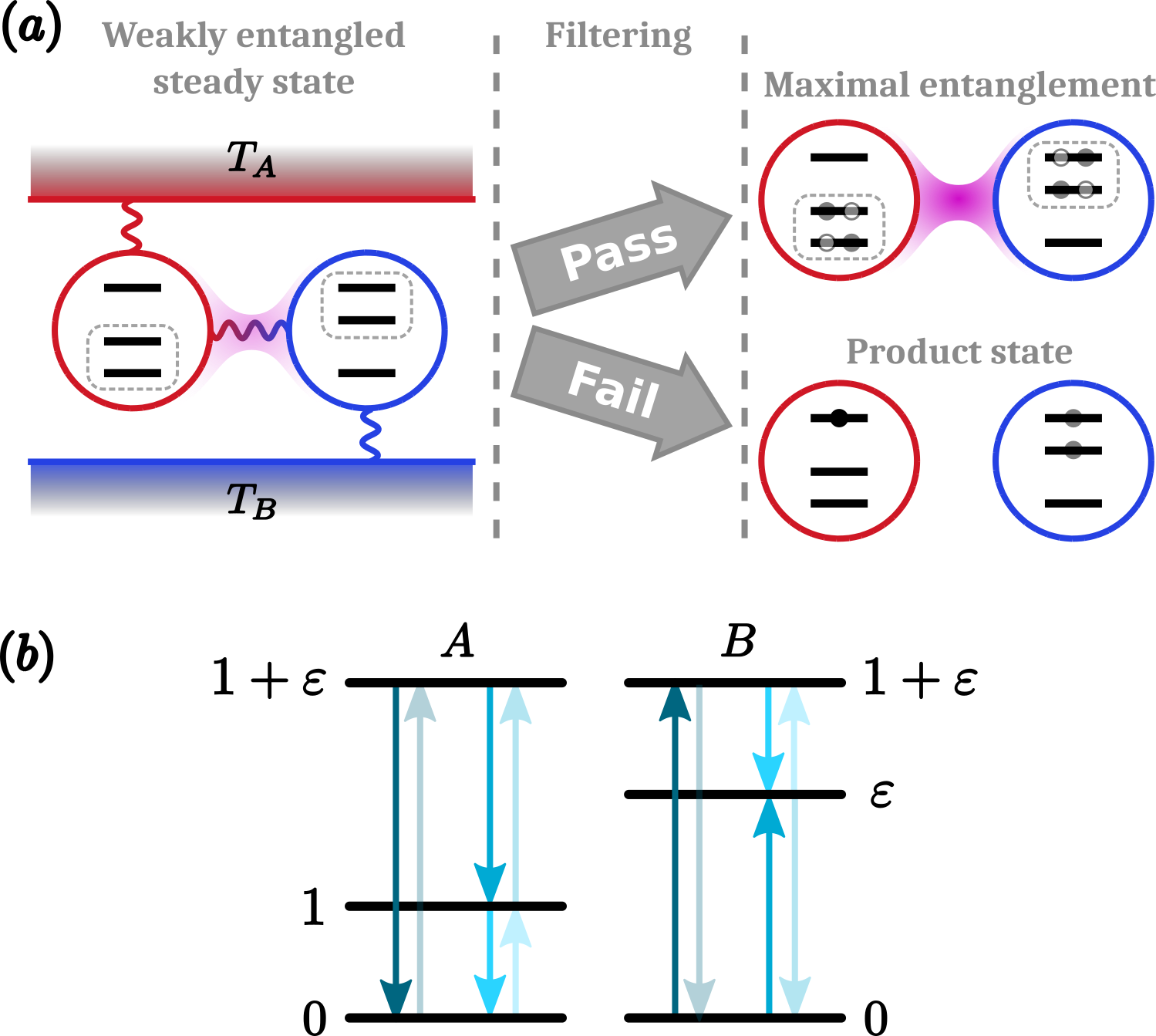}
\caption{\textbf{(a)} Qutrit thermal machine. Two qutrits are coupled to each other and to hot and cold thermal baths. The interaction with the baths drives the qutrits into a steady state featuring weak entanglement. Local filters project onto qubit subspaces on each side (dashed boxes). Upon success, the system is projected into a strongly entangled two-qubit state. Failure leaves the qutrits in a separable state, and the process must be restarted. \textbf{(b)} Level structure for the two qutrits. Arrows indicate the transitions involved in the interaction Hamiltonian.}
\label{fig.qutritscheme}
\end{center}
\end{figure}

On the other hand, it was shown that steady-state entanglement can be obtained in systems out of thermal equilibrium. This was first discussed for an atom coupled to two cavities driven by incoherent light \cite{Plenio2002}, and later for many-body systems \cite{Hartmann2006,Hartmann2007}, interacting spins \cite{Quiroga2007,Znidaric2012}, atoms in a thermal environment \cite{Bellomo2013a,Bellomo2013b}, and mechanical oscillators \cite{Boyanovsky2017}. In this context, Ref.~\cite{Brask2015} discussed what is arguably the simplest setting, namely a two-qubit system, where one qubit is connected to a hot bath and the other to a cold bath. This setup is promising for implementations in superconducting systems and quantum dots. Overall, the out-of-equilibrium approach thus opens interesting perspectives for dissipative entanglement generation. However, its main drawback so far is the fact that the generated entanglement is typically very weak, and thus not directly useful for applications. 

Here we offer a solution to this problem, presenting a scheme in which maximal entanglement can be generated in a heralded manner, through incoherent interactions with thermal baths alone. Specifically, a pair of $(d+1)$-dimensional systems is first driven to an entangled steady state, from which maximal entanglement in dimension $d$ can then be heralded via local filters. The procedure is implemented by a simple quantum thermal machine, operating out of equilibrium between two heat baths at different temperatures. Moreover, for $d=2,3$ we prove that any pure entangled state can be obtained without additional filtering, indicating that this holds for any $d$. Finally, we discuss experimental prospects considering an implementation in superconducting systems.

\section{Two-qutrit thermal machine}
\label{sec.twoqutritmachine}
The setup we consider is illustrated in \figref{fig.qutritscheme}(a). Two three-level systems (i.e.~qutrits) interact with each other, and independently with two thermal baths at different temperatures $T_A$ and $T_B$ (in the following, $T_A > T_B$ will be the relevant setting for entanglement generation). This out-of-equilibrium situation drives the two-qutrit system into a steady state, which is weakly entangled. A local filter is then applied to each qutrit, projecting the system onto a two-qubit subspace (as indicated by the dashed boxes). If the filter succeeds, the final state is arbitrarily close to a target two-qubit state. This target state can be any pure, entangled state, and in particular may be maximally entangled. If the filter fails, the system is left in a product state with no entanglement, and the process is restarted.

Each qutrit is described by a Hamiltonian $H_A$, $H_B$, and their interaction by $H_{int}$. We take the energy level structure illustrated in \figref{fig.qutritscheme}(b)
\begin{align}
H_A & = (\ket{1}_A\bra{1} + (1+\varepsilon)\ket{2}_A\bra{2})\otimes\openone_B , \\
H_B & = \openone_A\otimes(\varepsilon\ket{1}_B\bra{1} + (1+\varepsilon)\ket{2}_B\bra{2}) ,
\end{align}
where, without loss of generality, we set the ground state energies to zero and the first gap of qutrit A to 1 (throughout the paper, we work in units where $\hbar=k_B=1$). We are interested in autonomous processes, which require no external work input. This means that $H_{int}$ must be time independent and preserve the total energy, i.e.~$[H_{int},H_A+H_B]=0$. There are three possible energy-preserving transitions. Hence, writing $\ket{ij} = \ket{i}_A\ket{j}_B$, the most general form of the interaction is
\begin{equation}
\label{eq.qutritHint}
\begin{split}
H_{int} = \, & g_1 \ket{02}\bra{20} + g_2 \ket{11}\bra{20} + g_3 \ket{11}\bra{02} \\ & +  h.c.,
\end{split}
\end{equation}
where $g_1$, $g_2$, and $g_3$ denote the interaction strengths.  

To enable a fully analytical treatment, we first describe the evolution of the system in contact with the thermal baths by a simple reset model \cite{Linden2010}. When considering potential implementations below, we confirm that our results hold also under a Lindblad-type description of the open system. The reset model leads to the following master equation \footnote{Note that we are using a local master equation, where the dissipation induced by each bath acts locally on the subsystem connected to that bath. Recent works, analysing thermal machines similar to those employed here, have shown such a local approach to provide very good agreement with the exact dynamics for weak coupling, which is the regime of interest here \cite{Hofer2017,Onam2017}.}
\begin{equation}
\label{eq.master}
\begin{split}
\frac{\partial \rho}{\partial t}=i[\rho,H] & +p_A\left(\tau_A\otimes \Tr_A\rho-\rho\right) \\ & + p_B\left( \Tr_B\rho\otimes \tau_B-\rho\right) ,
\end{split}
\end{equation}
where $H=H_A+H_B+H_{int}$ is the total Hamiltonian, $p_A$, $p_B$ are coupling constants, and $\tau_A$, $\tau_B$ are thermal states, that is, $\tau_i= \exp(-H_i/T_i) / \Tr[\exp(-H_i/T_i)]$ for $i=A,B$. One can interpret \eqref{eq.master} as describing a process where, at each instance of time, each qutrit is either left unchanged or reset to a thermal state at the temperature of the bath, with resets happening at rates $p_A$, $p_B$. To ensure validity of our master equation, we always work in the perturbative regime where $g_1,g_2,g_3, p_A,p_B \ll 1,\varepsilon$.

To understand how the machine can generate entanglement, first note that the two-qubit subspace selected by the filters is spanned by the states $\{\ket{01}, \ket{12}, \ket{11}, \ket{12}\}$. Clearly, the transition $g_3$ generates coherence if the system is already in this subspace, between $\ket{11}$, and $\ket{02}$ thus creating entanglement. Interaction with the cold bath will tend to drive the cold qutrit towards the ground state, taking the system out of the filtered subspace. Transition $g_3$ cannot bring the system back, but transitions $g_1$ and $g_2$ do. In addition, the combination of these two transitions also generates entanglement because
\begin{equation}
\left[ H_{g1}, H_{g2} \right] = g_1 g_2 \left( \ket{02}\bra{11} - \ket{11} \bra{02}\right) \, ,
\end{equation}
where $H_{g1} = g_1  \ket{02}\bra{20}  + h.c.$ etc. If the cold bath temperature is low, the cold qutrit will tend to be in the ground state, and the system will only get excited into the filtered subspace whenever the joint state is $\ket{20}$. The interaction will then generate a pure, entangled state. Resets induced by the cold bath drive the system out of the filtered subspace and hence do not degrade the purity of the filtered state. Resets induced by the hot bath, on the other hand, do destroy coherence there, reducing the purity. Nevertheless, some hot resets are necessary to populate the state $\ket{20}$. We thus expect the best entanglement to be generated when $T_A$ is large, $T_B$ is close to zero, and $p_A \ll p_B$.

\begin{figure}
\centering
\includegraphics[width=0.95\linewidth]{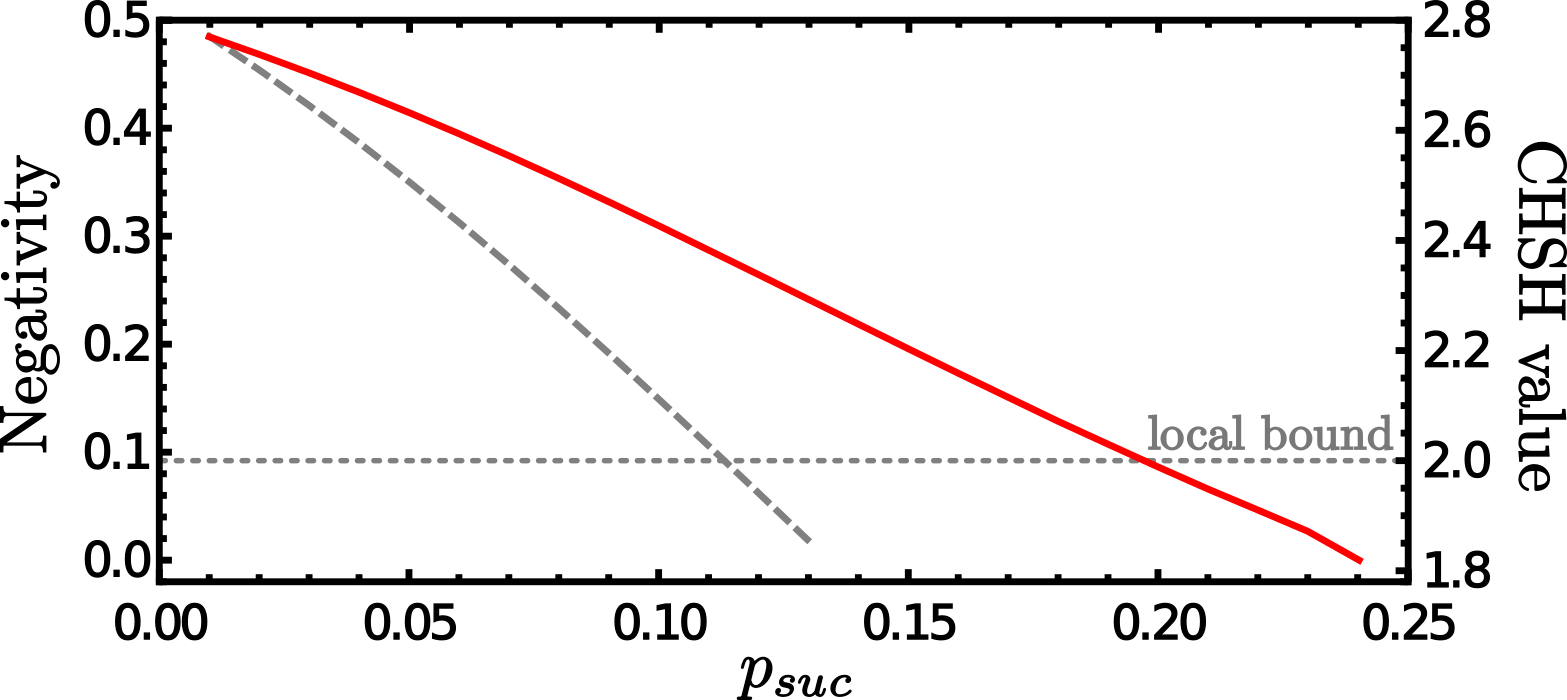}
\caption{Optimal negativity (solid, left axis) and CHSH value (dashed, right axis) vs.~postselection success probability. The dotted line shows the local bound above which the CHSH Bell inequality is violated.}
\label{fig.qubittradeoff}
\end{figure}

We have derived the steady-state solution $\bar{\rho}$ of \eqref{eq.master} in the limit of a maximal temperature gradient, $T_A\rightarrow \infty$, $T_B\rightarrow 0$ (see \appref{app.qutritsteady}). To obtain the final state, a local filter is applied to each qutrit, defined by projectors $\Pi_A=\ket{0}_A\bra{0}+\ket{1}_A\bra{1}$ and $\Pi_B=\ket{1}_B\bra{1}+\ket{2}_B\bra{2}$. The normalised, postselected state is
\begin{equation}
\label{eq.postselectedstate}
\rho' = \frac{1}{p_{suc}} (\Pi_A \otimes \Pi_B) \bar{\rho} (\Pi_A \otimes \Pi_B) ,
\end{equation}
where $p_{suc} = \Tr[(\Pi_A \otimes \Pi_B) \bar{\rho}]$ is the probability for the filtering to succeed. We take all the interaction strengths equal, $g_1=g_2=g_3=g$. In this case, the state after filtering becomes
\begin{equation}
\label{eq.filteredstate}
\rho' = \left(
\begin{array}{cccc}
 \frac{p_A}{4 p_A+6 p_B} & 0 & 0 & 0 \\
 0 & \frac{p_A+3 p_B}{4 p_A+6 p_B} & \frac{3 p_B}{4 p_A+6 p_B} & 0 \\
 0 & \frac{3 p_B}{4 p_A+6 p_B} & \frac{p_A+3 p_B}{4 p_A+6 p_B} & 0 \\
 0 & 0 & 0 & \frac{p_A}{4 p_A+6 p_B} \\
\end{array}
\right) .
\end{equation}
As expected, the highest purity of $\rho'$ is obtained when the ratio $\mu=p_A/p_B$ is small. For $\mu\rightarrow 0$, the state $\rho'$ tends to a pure, maximally entangled state (relabelling the basis states of the qubit subspace to $\ket{0}$, $\ket{1}$) 
\begin{equation}
\label{eq.targetqubitstate}
\ket{\psi_+} = \frac{1}{\sqrt{2}}(\ket{01}+\ket{10}). 
\end{equation}
Thus our machine can generate entanglement arbitrarily close to maximal. In addition, it is interesting to note that different choices for the interaction strengths enable the generation of other entangled states. Specifically, as shown in \appref{app.qutritsteady}, taking $g_1=g\cos(\theta)$, $g_2=g\sin(\theta)$, $g_3=0$ generates any partially entangled state of the form $\ket{\psi_\theta} = \sin(\theta)\ket{01}+\cos(\theta)\ket{10}$. We note that \eqref{eq.filteredstate} holds for any value of $g$. Hence the limit $\mu\rightarrow 0$ can be taken while keeping the ratio of $g/p_A$ fixed, retaining the validity of the local master equation.

\begin{figure}[!t]
\begin{center}
\includegraphics[width=0.95\linewidth]{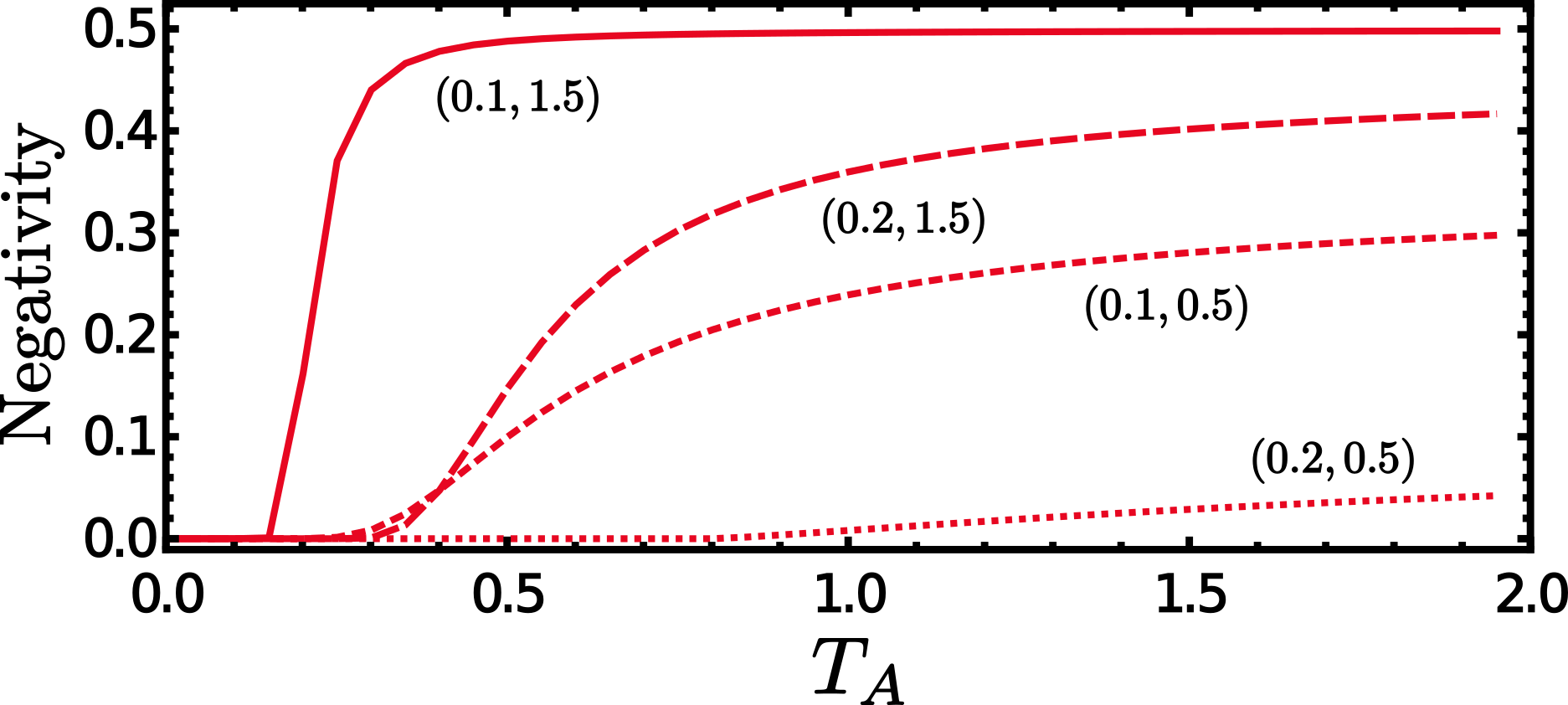}
\caption{Entanglement generation for finite temperatures. The numbers given for each curve are $(T_B,\epsilon)$ (in units of the first energy gap of qutrit A equal to 1). To make optimisation over the coupling parameters tractable, we maximise the off-diagonal element of the output state rather than the negativity directly. The curves therefore represent lower bounds.}
\label{fig.temperature}
\end{center}
\end{figure}

There is a trade-off between the probability for successful filtering and the quality of $\rho'$. The success probability tends to zero for both small $\mu$ (for fixed $g$) and small $g$ (for fixed $\mu$). In the two cases, respectively
\begin{equation}
\label{eq.qubitpsuc}
\begin{split}
p_{suc} & \approx \frac{1}{3}\frac{p_A}{p_B} , \hspace{0.3cm} \text{and} \\
p_{suc} & \approx \frac{2 (2p_A + 3p_B)}{9p_B (p_A+p_B)^2} g^2 .
\end{split}
\end{equation}
Adjusting the coupling parameters to increase $p_{suc}$ results in a final state $\rho'$ with a smaller overlap with the target pure state \eqref{eq.targetqubitstate}. Nevertheless, states of high quality can be generated. In \figref{fig.qubittradeoff} we show the maximal negativity \cite{Vidal2002} as well as the value of the Clause-Horne-Shimony-Holt (CHSH) quantity \cite{Clauser1969} for varying $p_{suc}$ (we optimise over $g$, $p_A$, and $p_B$, while imposing the perturbative regime). The negativity is an entanglement monotone ranging from 0 (separable) to 1/2 (maximally entangled) for qubits. Twice the negativity is a lower bound on the concurrence (which ranges from 0 to 1) \cite{Verstraete2001}. We see that $\rho'$ remains entangled up to $p_{suc} \approx 0.25$ and nonlocal up to $p_{suc} \approx 0.12$.

The machine thus provides a heralded source of entangled states: running the machine continuously, the system remains in the steady state until the entangled state is needed, at which point the filtering is performed. If filtering fails, the machine is allowed to return to the steady state, and another attempt can be made. A quasi-deterministic source can be constructed by running several machines in parallel. With $n$ machines, the probability for obtaining a successful projection in at least one of them scales as $1 - (1-p_{suc})^n$. Failure is exponentially suppressed in $n$.

In addition to the trade-off between success probability and quality of the postselected state, controlled by the coupling parameters, the temperatures also influence the generated entanglement. So far, we have taken a maximal temperature gradient, $T_A \rightarrow \infty$, $T_B \rightarrow 0$. In \figref{fig.temperature} we plot attainable negativity for finite temperatures. We see that, as might be expected, it is always better to take the hot bath temperature as large as possible, maximising the temperature gradient. As the cold bath temperature increases or the gap size $\varepsilon$ decreases, the hot bath temperature required to generate entanglement increases, and the maximal amount of attainable entanglement decreases. So, to maximise the entanglement, it is desirable to make $T_B$ small and $\epsilon$ large (note though, that $p_{suc}$ decreases with increasing $\varepsilon$).

\section{Two-qudit thermal machine}
\begin{figure}[!t]
\begin{center}
\includegraphics[width=0.65\linewidth]{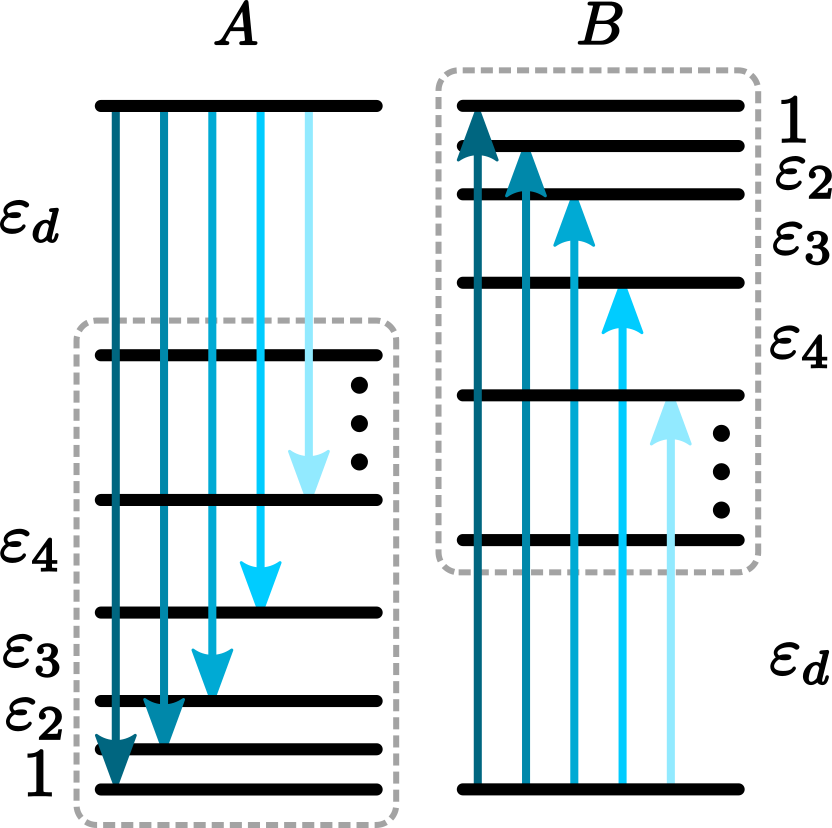}
\caption{Level structure of the two $(d+1)$-level systems in the qudit thermal machine. Arrows indicate the transitions involved in the interaction Hamiltonian. The dashed boxes indicate the $d$-dimensional subspaces to which the steady state is filtered to obtain the final state.}
\label{fig.quditlevels}
\end{center} 
\end{figure}

The scheme considered above can be generalised to create entangled states of two $d$-level systems, using a $(d+1)$-level thermal machine. The setup is the same as in \figref{fig.qutritscheme}(a), with the qutrits replaced by $(d+1)$-level systems, with level structures as illustrated in \figref{fig.quditlevels}. Denoting the energy gaps by $\varepsilon_k$ (with $\varepsilon_1=1)$, and setting $E_k^A = \sum_{l=1}^k \varepsilon_l$ and $E_k^B = \sum_{l=1}^k \varepsilon_{d-l+1}$, the free Hamiltonians are
\begin{equation}
\begin{split}
H_A & = \sum_{k=1}^d E_k^A \ket{k}_A\bra{k}\otimes\openone , \\
H_B & = \sum_{k=1}^d   \openone\otimes E_k^B\ket{k}_B\bra{k} ,
\end{split}
\end{equation}
and the interaction Hamiltonian is
\begin{equation}\label{Hqudit}
H_{int} = \sum_{k=1}^{d} g_k \ket{d,0}\bra{k-1,d-k+1} + h.c. ,
\end{equation}
corresponding to the transitions indicated on \figref{fig.quditlevels}. The evolution is again described by the master equation \eqref{eq.master}.

We will focus on the generation of a maximally entangled qudit state
\begin{equation}
\ket{S_d}=\frac{1}{\sqrt{d}}\sum_{k=1}^{d}\ket{k-1,d-k}.
\end{equation}
In that case, it suffices to set all the interaction strengths equal, $g_k = g/\sqrt{2}$ (the $\sqrt{2}$ ensures consistency with the qutrit case). In the limit $T_A\rightarrow\infty$, $T_B\rightarrow 0$, the steady state solution of \eqref{eq.master} can then be derived analytically for any value of $d$. It is given in \appref{app.quditsteady}. In analogy with the qutrit case, we consider local projections onto $d$-dimensional subsystems on each side, given by $\Pi_A = \openone-\ket{d}_A\bra{d}$ and $\Pi_B = \openone-\ket{0}_B\bra{0}$, and the state after successful filtering is again computed as in \eqref{eq.postselectedstate}. We find that, as before, high purity is attained when $p_A \ll p_B$, and the state tends to $\ket{S_d}$, as desired. Thus, our scheme is able to generate maximally entangled states in any dimension. The success probability is given by
\begin{equation}
\label{eq.quditpsuc}
p_{suc} = \frac{(d-1) g^2 p_A ((d-1) p_A+d p_B)}{d^2 \left(g^2 \xi + p_A p_B (p_A + p_B)^2\right)} ,
\end{equation}
where $\xi = \left(2 (d-1) p_A p_B+(d-1) p_B^2+p_A^2\right)$. One can check that this agrees with \eqref{eq.qubitpsuc} for $d=3$. Note that $p_{suc}$ scales like $1/d$ for large $d$, unless $g \sim 1/\sqrt{d}$. 

From $\ket{S_d}$, any pure two-qudit state can be obtained via biased filtering and local operations \cite{Nielsen1999}. However, given that any pure, entangled state of two qubits can be generated directly using the qutrit machine by adjusting the coupling strengths, it is natural to ask whether the same holds for qudits. In \appref{app.allpurequtrits}, we prove this for $d=3$, suggesting that it generalises to arbitrary $d$. Note that such direct generation can be advantageous in terms of success probability.

\section{Implementation}

\begin{figure}[t]
\begin{center}
\includegraphics[width=1\linewidth]{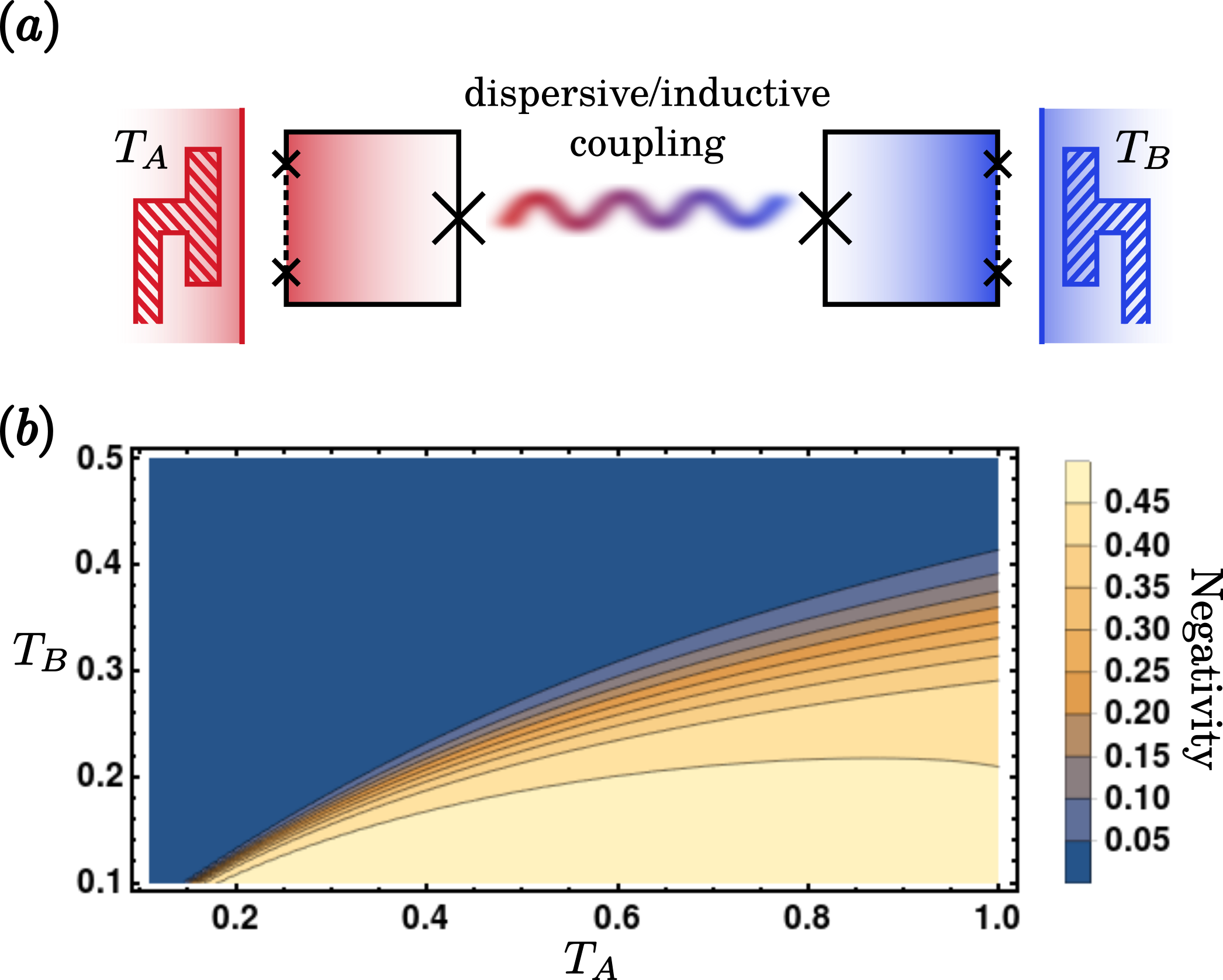}
\caption{\textbf{(a)} Implementation of the qutrit machine in circuit QED. Each fluxonium qutrit, depicted by their quantum circuit made of Josephson junctions \cite{Gu2017}, is capacitively coupled to a transmission line that plays the role of a thermal reservoir, see main text and Ref. \cite{Brask2015}. The flip-flop type interaction Hamiltonian between the two qutrits can be implemented either in the dispersive regime or by direct inductive coupling. \textbf{(b)} Negativity computed from the Lindblad model for $(\Gamma_A,\Gamma_B,\gamma,g,\epsilon)=(10^{-4},5\times 10^{-3},3.5\times 10^{-5},1.6\times 10^{-3},3)$ and $\Gamma_{B,12} = \Gamma_B/50$, see \appref{app.implementation} for the full Lindblad equation. All temperatures and energies are given in units of the first energy gap of qutrit A, taken to be 1 GHz. Near-maximal entanglement is generated in the bright region for experimentally relevant parameter values. } 
\label{fig.exp}
\end{center} 
\end{figure}

A variety of physical platforms might be considered for implementation of our scheme, including trapped atoms, ions, or solid-state artificial atoms. A promising platform is superconducting, circuit QED systems, which are generally good candidates for realizing quantum thermal machines \cite{Chen2012,Brask2015,Hofer2016a,Hofer2016b}. Here, we discuss prospects for a circuit QED implementation of the qutrit machine in more detail, and provide numerical evidence that strong entanglement generation can be achieved with parameter settings corresponding to state-of-the-art experimental capabilities, see Fig. \ref{fig.exp}.

Considering that the interaction \eqref{eq.qutritHint} requires the transition $\ket{0} \leftrightarrow \ket{2}$, fluxonium qutrits are good candidates for realizing the machine. In contrast to transmon qubits, for which selection rules forbid this transition, tuning of the magnetic quantum flux away of the sweet spot breaks quantum parity without inducing additional decoherence \cite{Manucharyan2009, Zhu2013}. Consequently, simple selection rules are absent and the transition $\ket{0} \leftrightarrow \ket{2}$ is allowed. Fluxomium artificial atoms have also recently shown outstanding performances in the context of quantum information processing thanks to their high tunability. In particular, their transition frequencies are in the range of hundreds of MHz to 30 GHz and the couplings to the baths can also be tuned from several kHZ to a few MHz \cite{Manucharyan2012, Pop2014, Lin2018}. In \cite{Lin2018}, it was even shown that complete decoupling from the environment is achievable.

With respect to the implementation of the interaction Hamiltonian, several coupling mechanisms are already available with fluxonium systems. First, similarly to transmon qubits \cite{Majer2007, Sillanpaa2007, DiCarlo2009}, fluxonium qutrits can be coupled capacitively or inductively via a cavity bus in the dispersive regime characterized by a strong frequency detuning of the qutrits and cavity with respect to their respective coupling strength to the cavity \cite{Manucharyan2012, Cottet}. Second, a possibly advantageous alternative is provided by a direct mutual inductive coupling as described in \cite{Chen2014} and proposed for fluxonium qutrits in \cite{Manucharyan2012}. Technicalities will depend on the actual frequencies that can be achieved experimentally.

Regarding the description of coupling mechanisms of each qutrit to a thermal bath, as well as the nature of the thermal baths in this setup, we refer to \cite{Brask2015}. It is also worth mentioning that fluxonium qutrits allow for flux-resolved spectroscopy, a technique to precisely determine all system frequencies \cite{Lin2018}.

Finally, the filtering procedure requires binary projective measurements onto a single energy level for each qutrit. That is, measurements which reveal whether or not the qutrit is in the corresponding state, but do not distinguish the remaining two states. This can be achieved by dispersive read-out in the regime where the dispersive shift is larger than the readout cavity line width (the photon-resolved regime) \cite{Schuster2007}. The shifts corresponding to each qutrit state will then be well separated and the transmittivity of the cavity at a frequency corresponding to, say, state $\ket{0}$ will be significant only when the qutrit is in this state, allowing for a binary projective measurement. A recent experiment operating in this regime was reported in \cite{Cottet2017}. Alternatively, two of the three shifts can be tuned to be identical. A binary projective measurement on qutrits using this technique was demonstrated in Ref.~\cite{Jerger2016}.

To model a circuit-QED implementation of the two-qutrit thermal machine and determine how much entanglement can be generated for reasonable parameter values, we use a master equation on standard Lindblad form. It describes dissipation due to coupling to bosonic baths, as well as pure dephasing, which is usually present in experiments. We note that it is possible to exactly map the reset model of \secref{sec.twoqutritmachine} to a Lindblad master equation of the form described here. This is discussed in \appref{app.implementation}. The equation (which replaces \eqref{eq.master}) can be written
\begin{equation}
\label{eq.Lindbladmaster}
\frac{\partial \rho}{\partial t} = i[\rho,H] + \mathcal{L}_A(\rho) + \mathcal{L}_A^z(\rho) + \mathcal{L}_B(\rho)  + \mathcal{L}_B^z(\rho) .
\end{equation}
Here, the dissipators $\mathcal{L}_A$ and $\mathcal{L}_B$ describe the effect of the thermal baths while $\mathcal{L}_A^z(\rho)$ and $\mathcal{L}_B^z(\rho)$ describe pure dephasing. We define
\begin{equation}
\mathcal{D}[O]\rho = O\rho O^\dagger - \frac{1}{2}\{O^\dagger O, \rho\} 
\end{equation}
to denote a standard Lindblad-type dissipator. Then
\begin{align}
\mathcal{L}_A(\rho) = \sum_{l=\pm} \sum_{k\in\{01, 12, 02\}} \Gamma_{A,k}^l \mathcal{D}[\sigma_k^l \otimes \openone] \rho  \, ,\\
\mathcal{L}_B(\rho) = \sum_{l=\pm} \sum_{k\in\{01, 12, 02\}} \Gamma_{B,k}^l \mathcal{D}[\openone \otimes \sigma_k^l] \rho \, ,
\end{align}
and
\begin{align}
\mathcal{L}_A^z(\rho) =  \sum_{k\in\{01, 12, 02\}} \gamma_{A,k} \,  \mathcal{D}[\sigma_k^z \otimes \openone] \rho \, ,\\
\mathcal{L}_B^z(\rho) =  \sum_{k\in\{01, 12, 02\}} \gamma_{B,k} \, \mathcal{D}[\openone \otimes \sigma_k^z] \rho \, .
\end{align}
Here, $\sigma_{mn}^\pm$ describe jumps between states $\ket{m}$ and $\ket{n}$ while $\sigma_{mn}^z$ describe phase flips between these states. Specifically,
\begin{equation}
\sigma_{mn}^+ = \ket{n}\bra{m} \, , \quad\quad \sigma_{mn}^- = \ket{m}\bra{n} \, ,
\end{equation}
and
\begin{equation}
\sigma_{mn}^z = \ket{m}\bra{m} - \ket{n}\bra{n} .
\end{equation}
The jump rates follow bosonic statistics ($j=A,B$)
\begin{align}
\Gamma_{j,mn}^+ & = \Gamma_{j,mn} \, n_B(\Delta E_{mn},T_j)  \, , \\
\Gamma_{j,mn}^- & = \Gamma_{j,mn} \, [1 + n_B(\Delta E_{mn},T_j)] \, .
\end{align}
In principle, the bath coupling constants $\Gamma_{A,k}$, $\Gamma_{B,k}$, and the pure dephasing rates $\gamma_{A,k}$, $\gamma_{B,k}$ could be different for each possible transition. For simplicity, here we take $\gamma_{A,k} = \gamma_{B,k} = \gamma$ to be the same for all transitions for both qutrits, and we take the bath couplings to be the same for all transitions $\Gamma_{A,k} = \Gamma_A$, $\Gamma_{B,k} = \Gamma_B$ with one exception. Jumps beteween states $\ket{1}_B$ and $\ket{2}_B$ of the cold qubit degrade coherence within the filtered subspace. Good entanglement generation therefore requires that $\Gamma_{B,12} < \Gamma_{B}$. This can be achieved by coupling through a bandpass filter centered away from the relevant transition frequency, reducing environmental damping for such transitions more strongly relative to jumps between the ground and excited states. The use of bandpass filtering to suppress environmental damping has been experimentally demonstrated \cite{Jeffrey2014,Bronn2015}.

We numerically solve \eqref{eq.Lindbladmaster} in the steady state and compute the amount of entanglement generate by our scheme.  Values for the different parameters (interaction strength $g$, qutrit energies $\epsilon$, bath coupling rates $\Gamma_A$, $\Gamma_B$, and pure dephasing rate $\gamma$) are taken from recent experimental achievements in circuit-QED architectures using fluxonium qutrits \cite{Manucharyan2012,Cottet2017,Kou2017}. The result is shown in \figref{fig.exp}(b). We see that near-maximal entanglement can be obtained. Thus, the scheme is a promising approach to demonstrating heralded entanglement using incoherent couplings to thermal baths. It is interesting to note in \figref{fig.exp}(b) that for fixed couplings, it is not optimal to maximise the temperature gradient. Maximal entanglement is obtained at a finite gradient.

\section{Conclusion}
We have demonstrated that combining incoherent couplings to thermal baths out of equilibrium with local filtering enables heralded generation of maximally entangled states in any dimension. The generated states can be made arbitrarily pure, at the price of lowering the filtering success probability. We have discussed an implementation of our scheme for qubit entanglement in superconducting systems, and found that prospects for a proof-of-principle experiment are good, with significant amounts of entanglement generated in the presence of decoherence and with limited temperature gradients. Interesting future perspectives include thermal generation of multipartite entanglement, and states useful for quantum computation or metrology.


We acknowledge helpful discussions with N.~Cottet and B.~Huard on implementations in superconducting systems. We acknowledge the Swiss National Science Foundation (Starting grant DIAQ, grant $200021\_169002$, and QSIT). GH acknowledges support from the Swiss National Science Foundation through the Marie-Heim V\"{o}gtlin grant no. 164466. MH acknowledges funding from the Swiss National Science Foundation (AMBIZIONE $PZ00P2\_161351$) and the Austrian Science Fund (FWF) through the START project Y879-N27.

\nocite{apsrev41Control}
\bibliographystyle{apsrev4-1}
\bibliography{revtexcontrol.bib,highdim_ent.bib}

\clearpage
\onecolumngrid
\appendix

\section*{Appendices}

In Secs.~\ref{app.qutritsteady} and \ref{app.quditsteady}, we provide details of the derivations steady-state solutions of the two-qutrit and two-qudit master equations. In \secref{app.allpurequtrits} we show how to generate any pure, entangled qutrit state. Finally, in \secref{app.implementation} we provide details of the implementation of our scheme in circuit QED.


\section{Finding the steady state and filtered state for the two-qutrit machine}
\label{app.qutritsteady}

Here, we explain how to derive the steady-state solution of the reset model master equation, Eq.~\eqref{eq.master} of the main text, for two qutrits, and how to obtain the filtered two-qubit state Eq.~\eqref{eq.filteredstate}.

The problem of finding an analytical solution is significantly simplified by the following observation: unless the interaction Hamiltonian induces transitions between $\ket{k,j}$ and $\ket{k',j'}$, there can be no coherence between these states in the steady state and the corresponding element in the density matrix vanishes, i.e., $\bra{k,j}\rho\ket{k',j'}=0$. This is because the dissipative processes locally reset each qutrit to a thermal state, which is diagonal, and hence do not generate any coherence. In the absence of the interaction Hamiltonian, dissipation would drive the system to a product of thermal states with no coherence. Note that, as seen in the previous section, in addition to the transitions directly present in $H_{int}$, it also induces second order transitions which need to be taken into account. Following such reasoning, one finds that there are only three non-zero off-diagonal elements in $\rho$. The density operator then takes the form:
\begin{equation}
\rho=\sum_{k,l=0}^{2}q_{kl}\ket{k,l}\bra{k,l}+ c_0 \ket{0,2}\bra{2,0} + c_1 \ket{1,1}\bra{2,0} + c_2\ket{0,2}\bra{1,1} + h.c
\end{equation}
where $q_{kl}$ are non-negative numbers that sum to one, and $c_0, c_1$ and $c_2$ are complex numbers which we can write as $c_k=v_k+iu_k$ with $v_k,u_k$ real. Plugging this ansatz into the master equation and requiring $\partial\rho/\partial t = 0$, we obtain three independent equations for the off-diagonals terms. Solving the real and imaginary parts of this equation system returns  $v_k$ and $u_k$ in terms of the $q_{kl}$. We are now faced with solving the system of equations corresponding to the diagonal of the right-hand-side of the master equation. This system of eight independent linear inhomogeneous equations can be written in the form $0=AX+W$ where $X=(q_{00},q_{01} \ldots, q_{21})^T$ and $A$ is a $8\times 8$ matrix depending on $g_1,g_2,g_3,p_A$ and $p_B$, and $W$ is a $8\times 1$ row-matrix accounting for the inhomogeneous part of the equation system. The solution can then be written $X=-A^{-1}W$. We note that, as the dissipation induced by resets leaves no subspace invariant, $A$ is always invertible when the rates $p_A$, $p_B$ are non-zero, and there exists a unique steady state. For maximal temperature gradient, $T_A\rightarrow\infty$, $T_B=0$, the solution can be computed analytically, although the expression is too unwieldy to display here.
	
To obtain the state given in the main text, one sets $g_1=g_2=g_3=g$. Applying the local filters to the steady state, as explained in the main text, and renormalising, one directly obtains Eq.~\eqref{eq.filteredstate}. Interestingly, the filtered state in this case is independent of $g$.

Interestingly, the scheme can also be adapted to generate any pure, entangled two-qubit state. This can be achieved by setting $g_3=0$ and taking $g_1=g\cos(\theta)$ and $g_2=g\sin(\theta)$. In this case, the filtered state becomes
\begin{equation}
\rho'= \begin{pmatrix}
r_1 & 0 & 0 & 0\\
0 & r_2 & t & 0\\
0 & t^* & r_3 & 0\\
0 & 0 & 0 & 1-r_1-r_2-r_3
\end{pmatrix},
\end{equation}
with
\begin{align}
& r_1=\frac{2 p_A \cos^2(\theta) \left(-g^2 p_A \cos(2 \theta )+g^2 (p_A+3 p_B)+3p_B(p_A+p_B)^2\right)}{(2 p_A+3 p_B) \left(-g^2 p_A \cos(4 \theta )+g^2 (p_A+6p_B)+6 p_B (p_A+p_B)^2\right)}\\
& r_2=\frac{2 \sin^2(\theta ) (p_A+3 p_B) \left(g^2 p_A \cos (2 \theta )+g^2 (p_A+3 p_B)+3 p_B (p_A+p_B)^2\right)}{(2 p_A+3 p_B) \left(-g^2 p_A \cos (4 \theta )+g^2 (p_A+6 p_B)+6 p_B (p_A+p_B)^2\right)}\\
& r_3=\frac{2 \cos^2(\theta ) (p_A+3 p_B) \left(-g^2 p_A \cos (2 \theta )+g^2 (p_A+3 p_B)+3 p_B (p_A+p_B)^2\right)}{(2 p_A+3 p_B) \left(-g^2 p_A \cos (4 \theta )+g^2 (p_A+6 p_B)+6 p_B (p_A+p_B)^2\right)}\\
& t=\frac{3 p_B \sin (2 \theta ) \left(g^2 (p_A+3 p_B)+3 p_B (p_A+p_B)^2\right)}{(2 p_A+3 p_B) \left(-g^2 p_A \cos (4 \theta )+g^2 (p_A+6 p_B)+6 p_B (p_A+p_B)^2\right)}.
\end{align}
To first order in the ratio $\mu=p_A/p_B$, when additionally $g \ll p_B$, one finds the simple expression
\begin{equation}
\label{eq.qubitpostselectedstate}
\rho' = \left(
\begin{array}{cccc}
\frac{1}{3}\mu c_\theta^2 & 0 & 0 & 0 \\
 0 & (1-\frac{1}{3}\mu) s_\theta^2 & (1-\frac{2}{3}\mu) c_\theta s_\theta & 0 \\
 0 & (1-\frac{2}{3}\mu) c_\theta s_\theta & (1-\frac{1}{3}\mu) c_\theta^2 & 0 \\
 0 & 0 & 0 & \frac{1}{3}\mu s_\theta^2 \\
\end{array}
\right) ,
\end{equation}
where $c_\theta=\cos(\theta)$, $s_\theta = \sin(\theta)$. For $\mu\rightarrow 0$, the state $\rho'$ thus tends to the pure state (relabelling the basis states of both qubit subspaces to $\ket{0}$, $\ket{1}$)
\begin{equation}
\label{eq.targetqubitstatepartial}
\ket{\psi_\theta} = \sin\theta\ket{0,1}+\cos\theta\ket{1,0}. 
\end{equation}
Hence, any pure, entangled two-qubit state can be obtained from the qutrit thermal machine (up to local unitaries). In particular, for $\theta=\pi/4$ (i.e.~$g_1 =g_2$), we again get a maximally entangled state.

The filtering success probability is given by
\begin{equation}
p_{suc} = \frac{2 g^2 p_A (2 p_A+3p_B) \left(g^2 p_A \cos(4\theta) - g^2 (p_A+6p_B) - 6p_B (p_A+p_B)^2\right)}{9A-9\left(B+C+D\right)} ,
\end{equation}
where
\begin{align}
A & = g^4 p_A \cos(4\theta) (p_A+p_B) (p_A+2p_B) , \\
B & = g^4 \left(p_A^3+11 p_A^2p_B+26 p_Ap_B^2+12p_B^3\right) , \\
C & = 2 g^2p_B (p_A+p_B)^2 \left(4 p_A^2+15 p_Ap_B+6p_B^2\right) , \\
D & = 6 p_Ap_B^2 (p_A+p_B)^4 .
\end{align}
We note that for small $p_A$ or $g$, the success probability depends only weakly on $\theta$.


\section{Finding the steady state and filtered state for the qudit machine}
\label{app.quditsteady}

In the following, we give the steady-state solution of the master equation, Eq.~(4) in the main text, for any $d\geq 3$, with the interaction Hamiltonian Eq.~(10). We work in the limit $T_A\rightarrow \infty$ and $T_B\rightarrow 0$. That is, we solve
\begin{equation}\label{app1}
	0=i[\rho,H_{int}]+p_A\left(\frac{\openone}{d+1}\otimes \Tr_A\rho-\rho\right)+p_B\left( \Tr_B\rho\otimes \ket{0}\bra{0}-\rho\right).
\end{equation}
Note that we have ignored the free Hamiltonian. We can do that since $\rho$ commutes with the free Hamiltonian in the steady state. This is because $H_{int}$ is energy preserving and can only generate coherence between states of the free Hamiltonians which are degenerate in energy (c.f.~the previous section).

We show that the following state solves \eqref{app1} for any $d\geq 3$.
\begin{align}\label{quditstate}
	&\rho_{d+1} =\frac{1}{N}\Bigg[\sum_{k,l=0}^{d}2g^2p_A^2\ket{k,l}\bra{k,l}+\sum_{k=0}^{d-1}c_1\ket{k,0}\bra{k,0}+c_2\ket{d,0}\bra{d,0}\\
	& + \sum_{k=0}^{d-1}2(d+1)g^2p_Ap_B\ket{k,d-k}\bra{k,d-k}  +\sum_{k=0}^{d-1}c_3\ket{d,0}\bra{k,d-k} + h.c \nonumber \\
	&+ \sum_{k=1}^{d-1}\sum_{l=1}^{d-k}2(d+1)g^2p_Ap_B \ket{k+l-1,d-k-l+1}\bra{k-1,d-k+1} + h.c\Bigg] , \nonumber
\end{align}
where we have defined coefficients
\begin{equation}
	\begin{split}
		& c_1=(d+1)p_Ap_B\left(p_A+p_B\right)^2+2g^2\left((d+1)^2p_B^2+2d(d+1)p_Ap_B\right)\\ 
		& c_2=p_A\left((d+1)p_B\left(p_A+p_B\right)^2+2(d+1)g^2dp_B\right)\\
		& c_3=i(d+1)gp_Ap_B(p_A+p_B)\\ 
		& N=(d+1)^2\left(p_Ap_B(p_A+p_B)^2+2g^2\left(p_A^2+2dp_Ap_B+dp_B^2\right)\right).
	\end{split}
\end{equation}

First we compute the following partial traces
\begin{eqnarray}
\Tr_A\left(\rho\right)=\frac{1}{N}\left[\sum_{l=0}^{d}2(d+1)g^2 p_A^2\ket{l}\bra{l}+\left(dc_1+c_2\right)\ket{0}\bra{0}+\sum_{k=0}^{d-1}2(d+1)g^2 p_Ap_B\ket{d-k}\bra{d-k}\right]\\
\Tr_B\left(\rho\right)=\frac{1}{N}\left[\sum_{k=0}^{d}2(d+1)g^2 p_A^2\ket{k}\bra{k}+\sum_{k=0}^{d-1}c_1\ket{k}\bra{k}+c_2\ket{d}\bra{d}+\sum_{k=0}^{d-1}2(d+1)g^2 p_Ap_B\ket{k}\bra{k}\right].
\end{eqnarray}
Subsequently, one can show that
\begin{multline}\label{Part2}
p_A\left(\frac{\openone}{d+1}\otimes \Tr_A\rho-\rho\right)+p_B\left( \Tr_B\rho\otimes \ket{0}\bra{0}-\rho\right)=\\
-\frac{1}{N}\Bigg[ -2d(d+1)g^2p_Ap_B(p_A+p_B) \ket{d,0}\bra{d,0}+2(d+1)g^2p_Ap_B(p_A+p_B) \sum_{k=0}^{d-1}\ket{k,d-k}\bra{k,d-k}\\
+c_3(p_A+p_B)\sum_{k=0}^{d-1}\ket{d,0}\bra{k,d-k}+c_3^*(p_A+p_B)\sum_{k=0}^{d-1}\ket{k,d-k}\bra{d,0}\\
+2(d+1)g^2p_Ap_B(p_A+p_B)\sum_{k=1}^{d-2}\sum_{l=1}^{d-1-k}\big(\ket{k+l-1,d-k-l}\bra{k-1,d-k}+\ket{k-1,d-k}\bra{k+l-1,d-k-l}\big).
\end{multline}
Similarly, extensive simplification of the commutator in \eqref{app1} gives
\begin{multline}\label{commutator}
[\rho,H_{int}]=\sum_{k=0}^{d-1}\Big(c_2g-2d(d+1)g^3 p_A p_B\Big)\ket{d,0}\bra{k,d-k}+2idg\text{Im}\left(c_3\right)\ket{d,0}\bra{d,0}\\
+\sum_{k=0}^{d-1}\Big(-c_2g+2d(d+1) g^3 p_A p_B\Big)\ket{k,d-k}\bra{d,0}-2ig\text{Im}\left(c_3\right)\sum_{\substack{k,l=0}}^{d-1}\ket{k,d-k}\bra{l,d-l}.
\end{multline}
Inserting \eqref{Part2} and \eqref{commutator} back into \eqref{app1}, the verification reduces to two equations
\begin{align}
&i\left(gc_2-2d(d+1) g^3 p_Ap_B\right)=c_3(p_A+p_B)
\\
&i\left(2dgi \text{Im}(c_3)\right)=-2d(d+1) g^2 p_Ap_B(p_A+p_B)
\end{align}
From the definition of $c_2$ and $c_3$, it is easily shown that both these equations are satisfied. Hence, the state \eqref{quditstate} is the steady-state of the thermal machine. 
	
Finally, we show that by applying suitable local filters to $\rho$, we obtain two maximally entangled $d$-level systems. The local projectors are
\begin{equation}
\Pi_A=\sum_{k=0}^{d-1}\ket{k}\bra{k} \hspace{15mm} \Pi_B=\sum_{l=1}^{d}\ket{l}\bra{l}. 
\end{equation}
The filtered state becomes
\begin{multline}
\rho'=\frac{\Pi_A\otimes \Pi_B \rho \Pi_A\otimes \Pi_B}{\Tr\left[\Pi_A\otimes \Pi_B \rho\right]}=\\
\frac{1}{\left(2g^2 p_A^2d^2+2d(d+1)g^2p_Ap_B\right) }\Big[\sum_{k=0}^{d-1}\sum_{l=1}^{d}2g^2p_A^2\ket{k,l}\bra{k,l}+\sum_{k=0}^{d-1}2(d+1)g^2p_Ap_B\ket{k,d-k}\bra{k,d-k}\\
\sum_{k=1}^{d-2}\sum_{l=1}^{d-k-1}2dg^2p_Ap_B \big(\ket{k+l-1,d-k-l}\bra{k-1,d-k}+\ket{k-1,d-k}\bra{k+l-1,d-k-l}\big)\Big].
\end{multline}
In the limit $p_A \ll p_B$ this indeed reduces to the maximally entangled state of two $d$-level systems
\begin{align}
\rho'=\ket{S_{d}}\bra{S_{d}} + O(\frac{p_A}{p_B}) . 
\end{align}


\section{Generating all pure, entangled states of two qutrits}
\label{app.allpurequtrits}

All pure entangled two-qutrit states can be written using the Schmidt-decomposition as
\begin{equation}
\ket{\psi^3_{\lambda_1,\lambda_2,\lambda_3}}=\sum_{i=0}^{2}\lambda_i\ket{i,i},
\end{equation}
with $\lambda\geq 0$ and $\lambda_0^2+\lambda_1^2+\lambda_2^2=1$. Here, we show that any such state can be generated using a two-ququart thermal machine and local filtering.

The machine consists of two ququarts (four-level systems) with an interaction Hamiltonian
\begin{equation}
H_{int}=g_0(\ket{0,3}\bra{3,0}+\ket{3,0}\bra{0,3})+g_1(\ket{1,2}\bra{3,0}+\ket{3,0}\bra{1,2})+g_2(\ket{2,1}\bra{3,0}+\ket{3,0}\bra{2,1}).
\end{equation}
Where we choose $g_i=g\lambda_i$ for some small constant $g$. In the limit of maximal thermal gradient, $T_A\rightarrow \infty$ and $T_B = 0$, the steady-state solution of the master equation can be derived using the method outlined in \secref{app.qutritsteady}. The steady state $\rho$ is then filtered to a space of two qutrits corresponding to the projectors $\Pi_A=\openone-\ket{3}\bra{3}$ and $\Pi_B=\openone-\ket{0}\bra{0}$. The filtered state $\rho'$ depends on $\lambda_0,\lambda_1,\lambda_2,p_A,p_B$ and $g$. We consider the limit in which $p_A \ll p_B$. This eliminates the dependence on $g$ and $p_B$. The resulting state is found to be
\begin{equation}
\rho' = \ket{\psi^3_{\lambda_0,\lambda_1,\lambda_2}}\bra{\psi^3_{\lambda_0,\lambda_1,\lambda_2}} + O(\frac{p_A}{p_B})  .
\end{equation} 
Thus, we can generate any pure entangled state of two qutrits.

Based on this result, and the corresponding case for qubits in the main text, we conjecture that any pure entangled state in any dimension can be generated by a generalisation of this thermal machine. Specifically

\subsubsection*{Conjecture}

Let the autonomous thermal machine of two $d+1$-level systems coupled to baths of temperature  $T_A\rightarrow \infty$ and $T_B = 0$ respectively, operate with an interaction Hamiltonian of the form
\begin{equation}
H_{int}=\sum_{k=0}^{d-1}g_k\ket{d,0}\bra{k,d-k}+h.c.
\end{equation}
where we take $g_i=g\lambda_i$ for some small constant $g$, and where $\{\lambda_i\}_i$ are the Schmidt coefficients of any pure entangled state of two systems of dimension $d$. Applying the projectors $\Pi_A=\openone-\ket{d}\bra{d}$ and $\Pi_B=\openone-\ket{0}\bra{0}$ to the steady-state of the system, and considering the limit $p_A \ll p_B$, the filtered state becomes
 \begin{equation}
 \ket{\psi^{d}_{\lambda_0,\ldots,\lambda_{d-1}}}=\sum_{i=0}^{d-1}\lambda_i\ket{i,i}.
 \end{equation}
In this work, we have shown this conjecture to be true for $d=2$ and $d=3$. In addition, we have checked numerically that the conjecture holds for $d=4$ and $d=5$ for 100 randomly chosen pure entangled states. Note that the number of adjustable paramters (the $g_k$) exactly match the number of Schmidt coefficients required to describe a pure state of two systems of dimension $d$.


\section{Reset vs Lindblad master equation}
\label{app.implementation}

The reset model considered in the main text is intuitive, amenable to analytical analysis, and captures the essential physics of a multipartite quantum system in contact with thermal baths. However, instantaneous thermal resets are a simplification with respect to realistic implementations. In this appendix, we first show that a reset master equation is exactly equivalent to a master equation on standard Lindblad form and derive an explicit mapping between the two. The corresponding Linblad master equation describes dissipation due to local coupling with bosonic thermal baths combined with additional pure dephasing. We then discuss how the optimal conditions for entanglement generation derived for the reset model translate to the Lindblad model.

\subsection{Equivalence for single qutrits}

Since a reset master equation generates Markovian (specifically semi-group) dynamics, there must exist a master equation of standard Gorini-Kossakowski-Sudarshan-Lindblad form which generates the same dynamics \cite{Gorini1976,Lindblad1976}. Here, we give an explicit mapping between these two forms.

We first consider a single qutrit and show that any reset master equation of the form
\begin{equation}
\label{eq.master1qutrit}
\frac{\partial \rho}{\partial t} = i[\rho,H] + \mathcal{L}_{res}(\rho) = i[\rho,H]+p \left(\tau-\rho\right)\,,
\end{equation}
where $p$ is a positive rate and $\tau$ is a thermal state, is equivalent to a master equation on standard Lindblad form given by
\begin{equation}
\label{eq.Lindblad1qutrit}
\frac{\partial \rho}{\partial t} = i[\rho,H] + \mathcal{L}_{lin}(\rho) = i[\rho,H] + \sum_{k\in\{01, 12, 02\}} \left( \Gamma_k^+ \mathcal{D}[\sigma_k^+] \rho + \Gamma_k^- \mathcal{D}[\sigma_k^-] \rho + \gamma_k \mathcal{D}[\sigma_k^{z}] \rho \right) \,.
\end{equation}
where the label $k$ runs over the three possible qubit subspaces of the qutrit, $\Gamma_k^\pm$ and $\gamma_k$ are positive rates, and $\sigma_k^{\pm}$ and $\sigma_k^z$ are jump operators acting on the qubit subspace labeled by $k$. Specifically
\begin{align}
\label{eq.jumpops}
\sigma_{mn}^+ = \ket{n}\bra{m} \, , \quad\quad \sigma_{mn}^- = \ket{m}\bra{n} \, , \quad\quad \sigma_{mn}^z = \ket{m}\bra{m} - \ket{n}\bra{n} .
\end{align}
The dissipators take the standard Lindblad form
\begin{equation}
\mathcal{D}[A]\rho = A\rho A^\dagger - \frac{1}{2}\{A^\dagger A, \rho\} .
\end{equation}
Having established a mapping between \eqref{eq.master1qutrit} and \eqref{eq.Lindblad1qutrit}, we generalise it to two coupled qubits below.

By a mapping between \eqref{eq.master1qutrit} and \eqref{eq.Lindblad1qutrit} we mean a set of relations defining $\Gamma_k^\pm$ and $\gamma_k$  in terms of $p$ and the elements of $\tau$ such that the right-hand sides of the two equations become equal. Since the Hamiltonian parts of \eqref{eq.master1qutrit} and \eqref{eq.Lindblad1qutrit} are the same, we only need to match the dissipators
\begin{equation}
 \mathcal{L}_{res}(\rho) = p \left(\tau-\rho\right) ,
\end{equation}
and
\begin{equation}
 \mathcal{L}_{lin}(\rho) = \sum_{k\in\{01, 12, 02\}} \left( \Gamma_k^+ \mathcal{D}[\sigma_k^+] \rho + \Gamma_k^- \mathcal{D}[\sigma_k^-] \rho + \gamma_k \mathcal{D}[\sigma_k^{z}] \rho \right) .
\end{equation}
The space of $3\times 3$ hermitian matrices is spanned by the projectors $\ket{m}\bra{m}$, $m=0,1,2$ and off-diagonals $\ket{m}\bra{n} + \ket{n}\bra{m}$ and $i\ket{m}\bra{n} - i\ket{n}\bra{m}$ with $m,n = 0,1,2$, $m<n$. The two dissipators will therefore act the same on any state $\rho$ if they act the same on each of these basis elements. By demanding $\mathcal{L}_{res}(\ket{m}\bra{m}) = \mathcal{L}_{lin}(\ket{m}\bra{m})$ we obtain a set of six equations (plus three redundant ones) which determine the $\Gamma_k^{\pm}$ in terms of $p$ and $\tau$. Similarly, by requiring $\mathcal{L}_{res}(\ket{m}\bra{n} + \ket{n}\bra{m}) = \mathcal{L}_{lin}(\ket{m}\bra{n} + \ket{n}\bra{m})$ for the three off-diagonals we obtain three more equations which determine the $\gamma_k$. Specifically, the solution is
\begin{align}
\label{eq.equivsol}
\begin{array}{lll}
\Gamma_{01}^- = p \tau_0 \,, & \quad \quad \Gamma_{01}^+ = p \tau_1 \,, & \quad \quad \gamma_{01} = \frac{1}{9} p (2-3\tau_2) \,,\\
\Gamma_{02}^- = p \tau_0 \,, &  \quad \quad \Gamma_{02}^+ = p \tau_2 \,, & \quad \quad \gamma_{02} = \frac{1}{9} p (2-3\tau_1) \,, \\
\Gamma_{12}^- = p \tau_1 \,, &  \quad \quad \Gamma_{12}^+ = p \tau_2 \,, & \quad \quad \gamma_{12} = \frac{1}{9} p (2-3\tau_0) \,.
\end{array}
\end{align}
where $\tau_0$, $\tau_1$, $\tau_2$ are the populations of the states $\ket{0}$, $\ket{1}$, $\ket{2}$ in the thermal state (i.e.~the diagonal elements of $\tau$). One can check that indeed using \eqref{eq.equivsol} one has $\mathcal{L}_{lin}(\rho) = \mathcal{L}_{res}(\rho)$ for any arbitrary qutrit state $\rho$. Explicitly, at a given temperature $T$, the populations are given by
\begin{equation}
\tau_m = \frac{e^{-E_m/T}}{\sum_{n=0}^2 e^{-E_n/T}} ,
\end{equation}
where $E_m$ is the energy of state $\ket{m}$, $m=0,1,2$. It follows that the jump rates in the Lindblad master equation satisfy detailed balance, as one would expect
\begin{equation}
\frac{\Gamma_{mn}^+}{\Gamma_{mn}^-} = e^{-(E_n-E_m)/T}
\end{equation}
We can then understand these jumps as being induced by a bosonic bath  \cite{Breuer, Gardiner, Schaller}
\begin{align}
\Gamma_{mn}^+ & = \Gamma_{mn} n_B(E_n-E_m,T) \, , \\
\Gamma_{mn}^- & = \Gamma_{mn} [1 + n_B(E_n-E_m,T)] \, , 
\end{align}
where
\begin{equation}
n_B(E,T) = \frac{1}{e^{E/T} - 1}
\end{equation}
is the Bose-Einstein distribution, and the coupling constant $\Gamma_{mn}$ for transitions between states $\ket{m}$ and $\ket{n}$ is given by
\begin{equation}
\label{eq.gammamap}
\Gamma_{mn} = p \frac{\tau_n}{n_B(E_n-E_m,T)} .
\end{equation}

\subsection{Equivalence for two qutrits}

The mapping derived between the single-qutrit master equations \eqref{eq.master1qutrit} and \eqref{eq.Lindblad1qutrit} can be applied directly to a system of two weakly coupled qutrits, as considered in the main text. Specifically, the reset master equation
\begin{equation}
\label{eq.master2qutrits}
\frac{\partial \rho}{\partial t} = i[\rho,H] + p_A (\tau_A \otimes \Tr_A(\rho) - \rho) + p_B (\Tr_B(\rho) \otimes \tau_B - \rho)
\end{equation}
is equivalent to the following local Lindblad master equation
\begin{align}
\label{eq.Lindblad2qutrits}
\frac{\partial \rho}{\partial t} = i[\rho,H] & + \sum_{k\in\{01, 12, 02\}} \left( \Gamma_{A,k}^+ \mathcal{D}[\sigma_{A,k}^+] \rho + \Gamma_{A,k}^- \mathcal{D}[\sigma_{A,k}^-] \rho + \gamma_{A,k} \mathcal{D}[\sigma_{A,k}^{z}] \rho \right) \nonumber \\
& + \sum_{k\in\{01, 12, 02\}} \left( \Gamma_{B,k}^+ \mathcal{D}[\sigma_{B,k}^+] \rho + \Gamma_{B,k}^- \mathcal{D}[\sigma_{B,k}^-] \rho + \gamma_{B,k} \mathcal{D}[\sigma_{B,k}^{z}] \rho \right) ,
\end{align}
where the jump operators are defined analogously to \eqref{eq.jumpops} above for each qutrit A and B locally. That is $\sigma_{A,k}^+ = \sigma_k^+ \otimes \openone$ and $\sigma_{B,k}^+ = \openone \otimes \sigma_k^+$, and similarly for the other jump operators. The mapping which makes the two master equations equivalent is given by \eqref{eq.equivsol} applied to each system A and B individually, as one can check.

Just as in the single-qutrit case, the jump rates in the Lindblad master equation correspond to bosonic baths. 
\begin{align}
\begin{array}{ll}
\Gamma_{A,mn}^+ = \Gamma_{A,mn} n_B(E^A_n-E^A_m,T_A) \, , \quad\quad &  \Gamma_{B,mn}^+ = \Gamma_{A,mn} n_B(E^B_n-E^B_m,T_B) \, , \\
\Gamma_{A,mn}^- = \Gamma_{A,mn} [1 + n_B(E^A_n-E^A_m,T_A)] \, , \quad\quad & \Gamma_{A,mn}^- = \Gamma_{B,mn} [1 + n_B(E^B_n-E^B_m,T_B)] .
\end{array}
\end{align}
When considering potential implementations of our scheme in the main text, we use a master equation of the form \eqref{eq.Lindblad2qutrits} for the numerical simulation, taking values for the bath coupling strengths $\Gamma_{A,mn}$, $\Gamma_{B,mn}$ and pure dephasing rates $\gamma_{A,mn}$, $\gamma_{B,mn}$ based on recent experimental works, as explained in the text.

\subsection{Optimal settings for generating maximal entanglement}

In the main text, we identified conditions under which our scheme generates a pure, maximally entangled state, using the reset model. Using the mapping above, we can translate these conditions to the Lindblad model.

The ideal temperatures for entanglement generation in the reset model are $T_A \rightarrow \infty$ and $T_B \rightarrow 0$. This means that the thermal populations become $\tau^A_0 = \tau^A_1 = \tau^A_2 = 1/3$ and $\tau^B_0=1$, $\tau^B_1 = \tau^B_2 = 0$. In turn, for the Lindblad jump rates, using \eqref{eq.equivsol} this implies that
\begin{equation}
\Gamma_{A,mn}^+ = \Gamma_{A,mn}^- , 
\end{equation}
and 
\begin{equation}
\Gamma_{B,01}^- = \Gamma_{B,02}^- \, , \quad\quad \Gamma_{B,12}^- = \Gamma_{B,mn}^+ = 0 .
\end{equation}
The former condition is satisfied in the Lindblad model also in the limit $T_A \rightarrow \infty$ since then $n_B(E^A_n-E^A_m,T_A) 	 \gg 1$. The latter condition can be satified in the limit $T_B \rightarrow 0$, where $n_B(E^B_n-E^B_m,T_B) \rightarrow 0$, if the coupling strength $\Gamma_{B,12}$ also vanishes.

Thus, we see that the Lindblad model is in principle compatible with the ideal limit for entanglement generation identified using the reset model, and one can thus expect entanglement generation to be possible also under such a more realistic model. We stress that it is not necessary to go to the ideal limit to achieve near-perfect entanglement generation. As shown in \figref{fig.exp} in the main text, using parameter values which are reasonable in the context of the current experimental state of the art, entanglement close to maximal can be attained.

\end{document}